# *The Shadow*: Coevolution Processes Between a Director, Actors and Avatars

Georges Gagneré
INREV-AIAC
Paris 8 University
Saint-Denis, France
georges.gagnere@univ-paris8.fr

## ABSTRACT

Andersen's tale *The Shadow* offers a theatrical situation confronting a Scholar to his Shadow. I program specific creatures that I called shadow avatar to stage the story with five of them and a physical narrator. Echoing Edmond Couchot's ideas about virtual people helping human beings to adapt to technological evolutions, I describe dynamics of coevolution characterizing the relationship between a director, actors, and shadow avatars during the process of staging *The Shadow*.

• Applied computing → Arts and humanities → performing arts

• Computing methodologies → Computer graphics → Animation → Motion capture

• Computing methodologies → Computer graphics → Graphics systems and interfaces → Mixed / augmented reality

## 1 Introduction

The show *The Shadow* [16] (figure 1) proposes to discover virtual actors, that I call shadow avatar, living in a virtual theater, through their encounter with a physical actor. These creatures have the particularity of sharing their existence between two spaces thanks to the possibilities of virtual reality simulations [23]. The shadow avatar thus exists in a 2D space like the natural shadow produced by any person illuminated by a point light source. But this shadow existing on a plane can also detach itself from its support to move like a flat silhouette in a 3D space (figure 2). It thus gives rise to a body of the shadow, similar to that which the Devil detaches from the hero of *Peter Schlemihl's Miraculous Story*, written by the romantic author Adelbert von Chamisso in 1814 [6].

Section 2 describes how the theme of the show has given rise to the notion of autonomy for virtual characters. This notion will be related to the intuitions of artist-theorist Edmond Couchot (1932-2020), who proposes a dynamic of coevolution between natural and virtual humans. Section 3 describes the technological context of avatar manipulation in which the show was created. Section 4 details the evolution of the tools made necessary to meet the artistic intentions, and the impact of this process on my way of directing actors and avatars. Section 5 discusses the influence of using the new tools on the actors' performance in relation to the avatars at two moments of the creation process: upstream, when recording the animations that are then used to write the visual part, and during the show's rehearsals to achieve its final form. Section 6 concludes with a discussion of the results and a perspective on further research.

## 2 Theatrical Context

The show is the occasion of a collaboration between a physical actor and five shadow avatars to stage the misadventures of a Scholar with his own shadow, written in 1847 by Hans Christian Andersen in the tale *The Shadow* [2].

### 2.1 The Story

During a stay in a hot country, a Scholar takes advantage of the cool evenings on his balcony while listening to be-witching music, played by a mysterious young woman living in the house opposite. One evening, seeing his own shadow projected on the house, he jokingly invites it to come inside to discover its secret.

The shadow disappears and does not return. As there is a lot of sun, a new shadow grows at the feet of the Scholar, who then returns to his cold country.

Several years pass when the former Shadow shows up one evening to redeem his freedom. Surprised, the Scholar graciously grants it to him, but he wants to know more about what happened in the mysterious house on the evening of the disappearance. The Shadow then recounts his extraordinary encounter with Poetry, how he had to fend for himself after being abandoned, and how he be-came very rich by sneaking into people's homes to listen to their unmentionable secrets and tell them pay for his silence.

After his departure, the Scholar continues to write his works on the True, the Beautiful and the Good, but he does not have much success and withers away. The Shadow comes back and convinces the Scholar to accompany him free of charge to thermal baths to recover his health. When they arrive at the baths, the Shadow meets a Princess who lets herself be seduced by his intelligence, so much so that she plans to marry him so that her kingdom can benefit from such wisdom. The Shadow making the Scholar pass for his own shadow, the latter refuses and tries to prevent the marriage from happening…

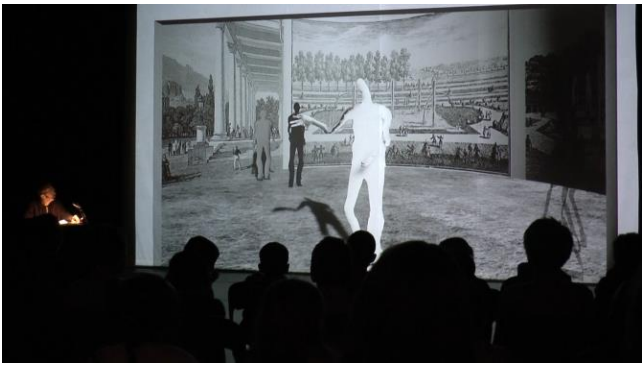

**Figure 1:** *The Shadow*: **encounter between the black Shadow and the white Princess in front of the grey Scholar**

## 2.2 A Processus of Autonomy

Andersen's character reaches a form of humanity after having been initiated by Poetry herself into the fundamental secrets of nature, formulated in poems of incandescent beauty that would burn any human being, but which instead humanized the Shadow.

> Do you know who lived in the house across the street from us? She was the most lovely of all creatures - she was Poetry herself. I lived there for three weeks, and it was as if I had lived there three thousand years, reading all that has ever been written. [...]Then I was in the anteroom. [...] That blaze of lights would have struck me dead had I gone as far as the room where the maiden was, but I was careful - I took my time, as one should. […] Had you come over, it would not have made a man of you, as it did of me. Also, I learned to understand my inner self, what is born in me, and the relationship between me and Poetry. Yes, when I was with you I did not think of such things, but you must remember how wonderfully I always expanded at sunrise and sunset. And in the moonlight I almost seemed more real than you. Then I did not understand myself, but in that anteroom I came to know my true nature. I was a man! [2]

Which transcendence precisely represents this terribly fascinating Poetry that the Scientist did not have the strength or the audacity to meet by himself? A quintessence of human knowledge or even the mysterious vitality from which all would proceed? Whether they are humans or shadows, the children of this lightning splendor would strive to coexist in dynamic ways that Darwin formalized in a theory of evolution that gives prominence to natural selection and merely skims over the principles of coevolution. In Andersen's Darwinist perspective, the original Shadow of the Scholar was boosted by Poetry and the rule of natural selection leads him to eliminate the Scholar.

## 2.3 Meeting Shadow Avatars

I chose to put this battle between the Scholar and his Shadow into perspective before two assemblies of spectators: that of physical humans and that of a people of shadow avatars which would inhabit a virtual theater placed in front of the spectators and whose construction is precisely described in [23]. The idea was to place ourselves on neutral ground and imagine the reaction of shadow avatars who would also be emancipated from their owners, and who could exist both as shadows on a surface, and as 3D shadows approaching the conditions of human existence. The shadow avatars, being able by nature to play a shadowless and emancipated shadow, take an empathetic attitude towards the Shadow character who returns to torment his original owner after being cowardly abandoned.

What do they think of this character who highlights the negative aspects of human beings and twists them to his advantage? From one point of view, a human being faces an autonomous double of himself that tragically escapes him. From another one, the human weaknesses of the Scholar and the flagrant ineffectiveness of his writings on "the True, the Beautiful and the Good" stand out with disturbing sharpness. Human society appears as an assembly of egoist people who consider the nature and all the other creatures as their slaves.

I chose to place the narration of this tragic confrontation on a terrain where shadow avatars and humans cohabit *a priori* without rivalry. The show *The Shadow* relies on the dual assumption that spectators may first believe in the existence of another world populated by strange entities that strongly resemble "traditional" shadows in some of their characteristics. Second, these entities should be able to relate to humans, including actors, to tell stories together on a theatrical stage.

The show begins with several of these entities who wander on a 2D plane around their virtual space, adjoining the physical stage, in front of the spectators who are gradually settling into the room. When everybody finished settling down, an actor gets up from the room to join the stage, establishes a first contact with the shadow avatars, and comes to sit down at a small table left to the screen and tell the tale by Andersen (figure 1).

Progressively, the shadow avatars, more reckless than the physical audience who remains quietly seated watching the scene, get hooked on the narration and come from 2D to 3D (figure 2). Initially mere spectators, the shadow avatars begin to replay dialogue situations, vocally accompanied by the narrator who digitally modulates his voice in real time depending on whether he is interpreting the Scholar, the Shadow, or the Princess.

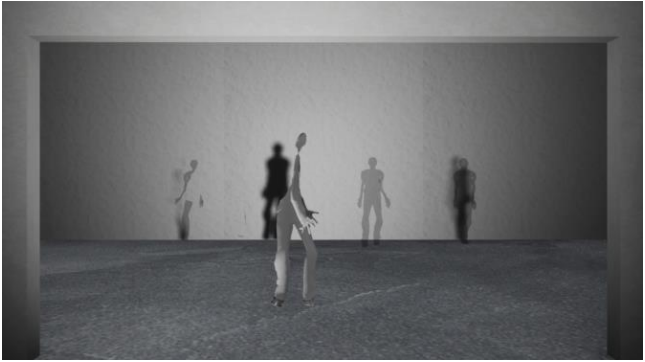

**Figure 2:** *The Shadow*: **Shadow avatars passing from 2D to 3D.**

During the show, the staging strives to make the public believe that the shadow avatars are autonomous and capable of interpreting the misadventures of the Scholar with his shadow. In practice, the shadow avatars are not autonomous entities endowed with their own life and therefore the question of their coexistence with humans is a simulacrum. The show is conceived in a fictional framework of the autonomy of a people of shadows in which Andersen's tale is embedded.

## 2.4 Couchot's Virtual People

This coexistence of two entities of a different nature is to be compared to an intuition of Edmond Couchot, artist and theoretician of digital art, to whom the Digital Image and Virtual Reality laboratory (INREV) of the Paris 8 University, of which he was the co-founder, paid tribute in November 2022 at the international conference "Pour un imaginaire numérique avec Edmond Couchot" (For a digital imaginary with Edmond Couchot). I recalled on this occasion the proposal he made of a coexistence between natural and artificial humans, among which he placed robots and avatars [18]. Taking note on the one hand of a shift to a postgenomic society freeing itself from the Darwinist determinism of natural selection thanks to the progress of biotechnologies and, on the other hand, of the continuation of the phenomenon of externalization of human cognition by increasingly complex processes of digital simulation including artificial intelligence research astonishing results, Couchot assumed the emergence of a virtual people constituting a subspecies of the human species.

> Artificially endowed with traits that are as close as possible to those of the natural human species, the virtual people would constitute a new species, or rather a sub-species, born of the intersection of research into post-genomic living and the externalization of cognition. The situation is reminiscent of that evolutionary episode when Neanderthals came face to face with Cro-Magnons, who also belonged to the Homo sapiens species but had a slightly different anatomy and culture. This encounter was followed by the rapid disappearance of the first subspecies, a fact that prehistorians have yet to fully explain. I won't propose a catastrophic scenario of the kind often evoked by science-fiction writers (the extinction of one subspecies of natural humans to the advantage of the other), but I will put forward a reasonable hypothesis: the probable effect of this confrontation would be to modify the natural human's physicality and encourage him to readapt to his new technological, societal and cultural environment, by providing him with the means to do so, just as any living species is obliged to mutate to adapt to its environment when it changes. [8]

Couchot proposed that the confrontation between the two species participated in the endless process of hominization, even mutation in the context of contemporary technological developments. According to him, performing arts were a favorable ground for this evolution by stimulating spectators' aesthetic attention and an acculturation to new emerging corporeities [9].

The creation of the show led to the evolution of artistic practices on three levels, allowing to figure out dynamics of coevolution between natural and artificial beings:

- the implementation of a real-time avatar animation framework and its evolution towards a specific arrangement of animations implying a new way of acting,
- the process of recording animations with a physical actor,
- the acting by the narrator.

## 3 Directing Avatars on a Mixed Stage

### 3.1 AvatarStaging

The staging of *The Shadow* is based on the possibility of directing avatars in 3D space. It is realized in a framework I call AvatarStaging [20], which combines devices and software tools to transpose movements performed by an actor equipped with a motion-capture device—the "mocaptor"—onto the avatars inside the virtual theater, which represents the virtual component of the mixed-reality stage.

AvatarStaging brings together three operations that let a mocaptor control an avatar and develop a relationship between said avatar and a physical actor on a mixed-reality stage. Figure 3 is a schematic representation of how these three operations work for a mocaptor, an avatar, and an actor, variations of which can be inferred in more complex configurations. In the first operation is capturing the motion itself. The mocaptor wears a full-body suit, which can be optical or inertial, and which makes it possible to collect data relating to the position of the sensors situated on specific areas of the mocaptor's body (Figure 3, zone C). The motion-capture data are transferred to a dedicated software where a virtual skeleton is inferred from the sensors positions. The movements of this skeleton are hence the result of a mathematical approximation and do not match exactly the mocaptor's actual movements. Moreover, multiple parasitic factors affecting data capture can cause distortions. Their impact on the artistic quality of the movements must constantly be assessed and corrected, if necessary.

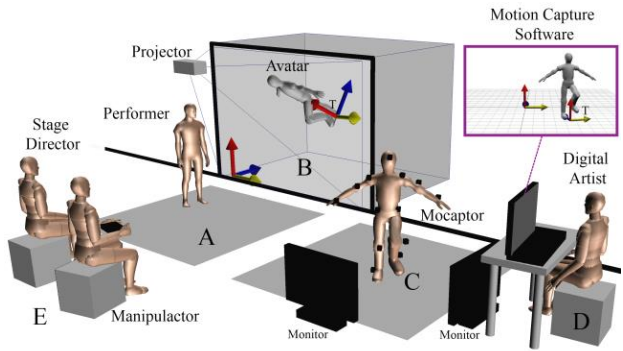

**Figure 3: AvatarStaging setup**

The second operation, called motion retargeting, involves transposing the movements of the first virtual skeleton onto the virtual skeleton of the avatar chosen to perform in the 3D space in dialogue with the actor. In *The Shadow*, the avatar has a very specific build, inevitably distorting the mocaptor's original motions, an aspect which I will discuss at greater length in relation to the project's artistic outcomes (see 5.1).

The third and last operation consists in organizing the positions of the avatar and/or its prerecorded animations according to the digital setting and to the positions of the physical actor (Fig. 3, zone B), using the AKN_Regie plugin I'll will introduce in the next section.

Consequently, the mocaptor present in space C develops a stage relationship to the actor on stage A thanks to the video feedback monitors placed around him, which enable him to see the result of his movements projected onto his avatar in digital space B. This dynamic is made possible by the three operations of motion capture, motion retargeting, and motion control in AKN_Regie.

### 3.2 Genesis of AKN_Regie

AKN_Regie is the result of my artistic evolution between 2015 and 2018 in contact with the progressive exploration of avatar staging that led to formalizing AvatarStaging [22]. My first contact with avatar manipulation took place in 2015 during a research-creation project at the Bordeaux Computer Science Laboratory dedicated to the use of a time sequencer under development, I-score [1] during the Virage project [3], which has since become Ossia Score [28], to control artistic media. At the time, I was working with Cédric Plessiet, a digital artist, who had created a project for me in the Unity3D game engine, enabling me to trigger animations in place for 3 shadow avatars. The experiment worked, but I soon found myself limited by the need to modify the avatars' control parameters [14]. With a view to a second residency the following year, I decided to familiarize myself with Unreal Engine, the video game engine developed by Epic Games [12], to build myself the finite-state machine used to sequence the animations, whose triggers were always under the control of the I-score sequencer [15].

Until then, I'd never directly confronted software programming to stage creations involving physical actors and interactive real-time video. I used to collaborate with artists or technicians who implemented the digital parts of the scenarios I designed and which we adjusted collectively [13]. The experience of 2016 made me realize that the ambition to direct avatars implied an in-depth understanding of how to program them.

Different experiences over 2017-18 led me to formalize AvatarStaging and develop AKN_Regie as an Unreal Engine plugin to perform the following operations with avatars controlled in real time by mocaptors:

- Retrieve data from real-time motion capture devices and apply it to avatars
- Precisely place avatars in 3D space and orient them with game controllers
- Program a sequence of successive configurations for manipulating avatars in 3D space
- Use the keyboard and MIDI controllers to manipulate the pipeline during a performance

[17] gives an idea of the architecture of the tool, which is based on visual programming using the Blueprint language specifically developed by Epic Games in Unreal Engine to facilitate use of the engine by people who don't know how to code C++ in written script form. It has been used as an authoring tool on various projects [32] [33] and has an instruction manual accessible to non-computer specialists on a dedicated website [11].

AvatarStaging and AKN_Regie enable an avatar to be brought to life on a mixed stage, both in real time and off-line when replaying a recording. The challenge posed by *The Shadow* project was to make the narrator and several shadow avatars cohabit in a single staging.

## 4 Evolution of Staging Approach

### 4.1 New Way of Thinking about Direction

Right from the start of the experiments with the shadow avatars in 2015, the intuition was to work on the quality of the sequences of animations to create a living relationship between the narrator and the avatars. Based on the dynamics of avatar animation deployed in video games, both for the Player Character and the Non-Player Characters, I assumed that the narrator could trigger successive actions by the shadow avatars to gradually involve them in the story, either through more active listening, or by replaying the dialog passages, leading to a staging of the tale. One solution would have been to produce an animated film using movements pre-recorded with the mocaptor. But the amount of work involved would have been enormous, and moreover, having to follow the fixed temporality of a film didn't fit in with my approach to live narration from a theatrical perspective.

My artistic intention was to confront the narrator and the audience with the people of the shadow avatars. The relationship between natural and virtual humans implied the primordial condition of live performance, in my opinion, of allowing the narrator his acting freedom, and in particular his pacing. The shadow avatars have neither face nor mouth. They can only mime the actions being narrated. The narrator would play the voices, with the shadow avatars completing the role by mimicking the actions. He therefore had to keep control over the sequence of mimed actions.

I then assume that it was possible to decompose all scenic gestures into two categories [19]:

- salient: actions with salient gestures are characterized by a starting point and an ending point within a perceivable timeline
- non-salient: actions which do not contain significant gestures and take on a "listening" or "waiting" position just before or after a salient action (category also referred as idle)

For an actor on stage, a salient action corresponds to speaking text with or without gesture, while the other actors take on physical positions without salient gestures, as if listening or waiting, so as not to disturb the focus on the speaking character. Sometimes, salient actions can also impact or interrupt each other. But the art of stage direction is to organize these salient actions from the point of view of the spectators' perception and keep the narrative thread.

## 4.2 Programming Perspective

I then extended my Unreal Engine programming skills to develop this functionality in AKN_Regie tool, which up to now had only been used to control avatars in real time with mocaptors. Non-salient actions are found in video game development under the term idle animation (referred to as an idle) – when a virtual character assumes a low-activity position. This indicates presence, but without any significant expressive gestures, and when the animation end point is blended back to its beginning it forms an endless loop. A breath or a position of attentive listening can be an idle. An idle can be played in forward or in reverse. It is usually necessary to tweak the animation with an editing software as for instance Motion Builder (Autodesk) to get a smooth loop transition without jump.

I proposed another way to build an idle loop from a piece of idle movement: adding in real time the animation to itself in reverse play mode. In this way the loop is continuous, instantaneously, without requiring any time costing editing. An avatar may maintain an idle loop a long time, waiting to carry out a future salient action. A trained eye will detect the loop pattern after a certain time, which degrades the avatar presence effect. But idles are often applied to avatars that are not at the center of the stage action. We call presence effect the quality that triggers the willing suspension of disbelief of a spectator confronted to a fiction, in that case, the fictitious live of an avatar. The notion was proposed by Coleridge to explain how a reader, or a spectator believes in a fiction, although he knows that it is a fiction [7].

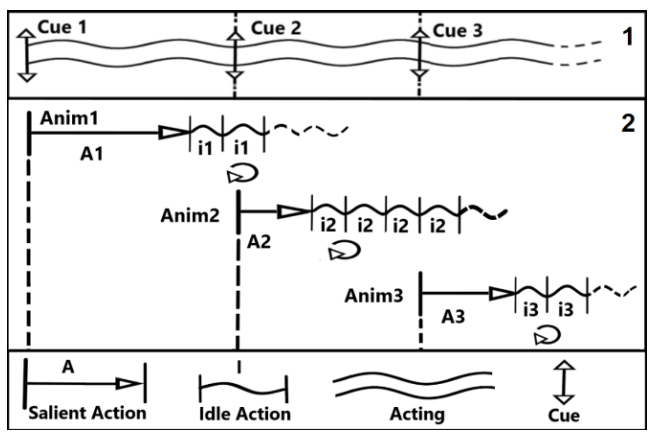

Figure 4: Principles of Salient-Idle gesture decomposition

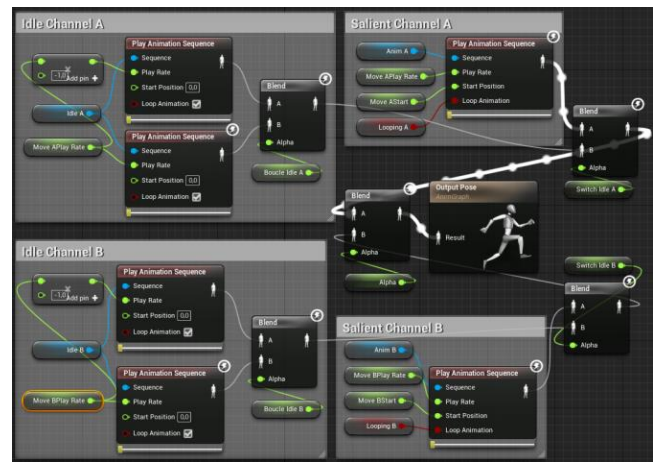

Figure 5: Blueprint part of Salient-Idle Player in AKN_Regie

This decomposition of each scenic movement was intended to let the narrator control the rhythm of the avatars' play, since each salient movement would end in a looped idle. The protocol for linking Salient-Idle animations according to cues that are not fixed in time is shown in figure 4. Cues 1, 2 and 3 are shown in panel 1, and are triggered at various flexible dates. In panel 2, each cue triggers a salient animation followed by a looping idle that allows the narrator to trigger subsequent cues at his own pace.

Figure 5 shows part of the blueprint programming that alternates Salient and Idle animations. [19] provides a more detailed description of this programming. Implemented under the name of Salient-Idle Player (SIP), the functionality enabled the construction of 50 minutes of continuous stage play for 5 characters, creating the illusion of their autonomy from the audience's point of view.

### 4.3 Unanticipated Results

An unanticipated result opened a real possibility of modifying the sequence of animations and deepening avatar direction. Figure 6 brings together three SetSalientIdle instructions to the shadow avatars playing the characters of the Scholar, the Shadow, and the Princess. Each SetSalientIdle has PlayRate and Transfer parameters that apply both on Salient and Idle Animations. It means that one can modify the speed of each part of the movement, and the blending time from Salient to Idle animation, and from forward Idle to backward same Idle animation in the looping operation. These parameters allow a fine tuning of the global movement pacing that has an impact on the avatar interpretation.

Moreover, as shown in figure 5, the principle of looping an idle is based on mixing in real time the animation with itself, played backwards as it ends. Indeed, if a Salient movement contains a piece of Idle one, it is possible to cut it in three parts (two shorter Salients and an in between Idle), and to use two SetSalientIdle functions to combine the original movement with a pause in between as long as necessary. It is then possible to suspend the course of the action and modify the parameters of each new animations to accentuate gesture scenic intentionalities.

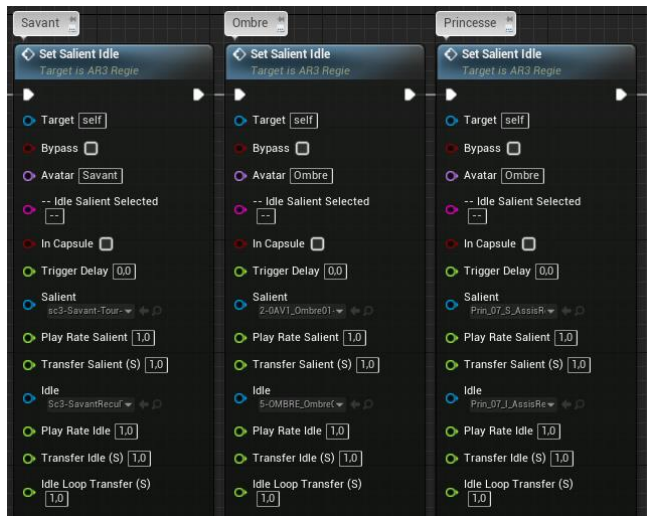

Figure 6: Salient-Idle Player parameters for three characters

These modifications are made on the fly, after the sequences have been cut directly in the game engine, the new functions implemented in the CueSheet, and the SetSalientIdle parameters tweaked appropriately. It gives the flexibility to adjust movement in real time, which is decisive when it comes to tweaking avatars presence effect in relationship with the narrator during rehearsals.

I find that the more in-depth programming of the animations that produced SIP allows me to rework the scenic intentions of the shadow avatars with a degree of autonomy from the original recordings. This enables an experimental exploration of the expressive dimension of gestures that would be difficult, if not impossible, to achieve with a physical actor. In return, it seems to me that the avatar direction work carried out with SIP enabled me to give more precise acting indications to the mocaptor when we recorded the animations. On the one hand, I was probably influenced by the subsequent use I was going to make of the movements, but it also seems to me that decomposing the movements allows me to better communicate the desired changes in intention.

## 5  Acting with Avatars

I will now describe how working with avatars has influenced the way the mocaptor and narrator play.

### 5.1  Inhabiting an Avatar with Empathy

The recording sessions were intense for the mocaptor because he had to organize his acting on the physical stage surrounded by feedback monitors to create convincing shadow avatar presence effect in the virtual theater (figure 7).

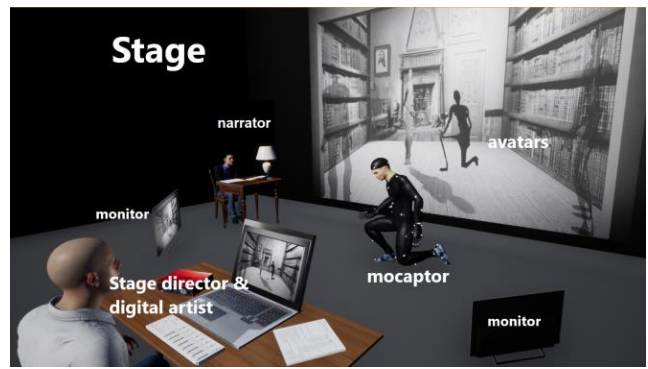

Figure 7: Rehearsals in on set previsualization context

The transmission of states of presence from the mocaptor to the avatar combines two processes in the deployment of scenic play [21]. First, there is the work of understanding the relationship to the avatar's body, which consists in adapting an inner proprioceptive awareness of the acting in relation to the avatar's movement potential. This concerns the direct relationship between the mocaptor and the avatar, which I have placed in a first circle of presence effect (cf. figure 8) and which corresponds to the work of appropriating the motion capture process in space C of figure 3.

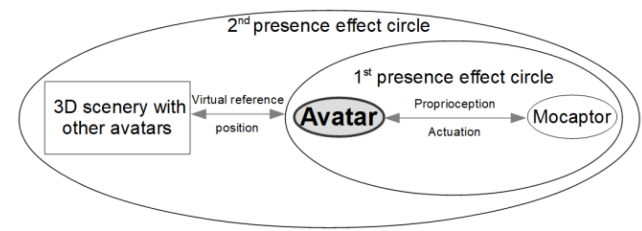

Figure 8: Presence effect circles during rehearsals

At the same time, the mocaptor must consider the avatar's relationship to its 3D environment, by memorizing virtual reference positions that are constructed with the director according to staging requirements and 3D scenography (cf. figure 8 - 2nd circle and space B in figure 3). He can also use the feedback monitors to check the accuracy of the avatar's scenic actions. The director plays an essential role in establishing virtual reference positions and the quality of presence in the second circle. But he can also help the mocaptor to make the avatar's control of the first circle more expressive. We can see that the second circle of presence invites the mocaptor to project itself in the avatar's place to give meaning to its actions in relation to the 3D environment, in a process of empathetic projection. Empathy can be defined as the ability to put oneself in another person's shoes, while maintaining one's own point of view [5].

During many of the rehearsals (figure 8), I worked simultaneously with the narrator playing the story while the mocaptor controlled the avatar on the same scenic setup used for the public performance (figure 10). It is an on-set previsualization setup that enhances the shadow avatars presence effect with the narrator during the rehearsals [29].

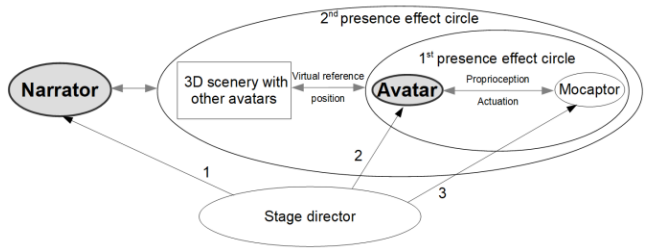

**Figure 9: Levels of direction during on-set previzualisation rehearsals : 1 towards Narrator, 2 Avatar and 3 Mocaptor**

In rehearsal time, I direct two entities of a different nature, the narrator and the shadow avatar, and my artistic goal is to establish relationship in between through the presence effect of the avatar. I know that the shadow avatar is animated by the mocaptor, but I must guide the three entities staying focused on the relationship between the narrator and the avatar to achieve a scenic co-presence of two beings of very different natures (arrows 1 and 2 in figure 9). This co-presence is what makes possible to immerse the spectator in the fiction. The shadow avatar has no autonomy, it is inhabited by the mocaptor in the first presence effect circle (arrow 3 in figure 9). All the animations are then processed with SIP (see section 4) to form a score that the narrator triggers cue by cue during the performance [19].

## 5.2 Taming Each Other

The narrator who plays with the shadow avatars also deploys empathy in the second part of the rehearsals where he is in relation with the combinations of animations previously recorded with the mocaptor. In a way, shadow avatars are partially independent on the bodily level. The alternance between salient and idle actions allows the narrator to keep control over the temporal unfolding of the performance. He triggers each salient action according to his acting, and the shadow avatars then wait for him to trigger the next one.

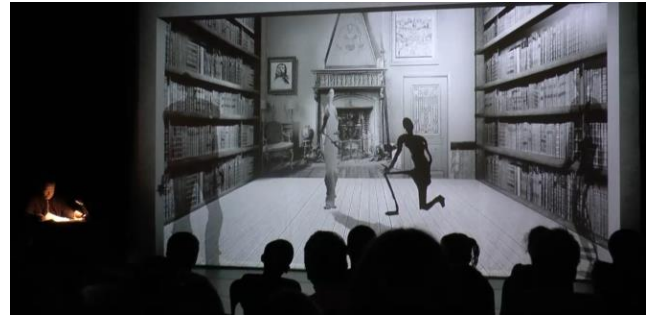

**Figure 10: Use of the recordings during the representation**

When the action is launched, the actor interprets the voice of the character that the shadow avatar embodies. The shadow avatars have no face or mouth to speak with. They all have the same body, with textures that differentiate them (for example in gray for the Scholar, in black for the Shadow or in white for the Princess, see figure 1). The actor plays with a head microphone that allows to transform his voice in real time according to the character played. The Shadow has half a dozen different lower voices, programmed with a specific patch based on Max-MSP (Cycling74) [10], to tune transformations according to different intentions. It helps the physical actor to vary the acting intentions in the low transformed pitch which sometimes tends to overwhelm the subtle variations of the natural voice. The actor finds the flexibility of interpretation which enables him to adapt to the movements of the shadow avatar he accompanies.

The mocaptor is no more present during this second part of the rehearsals. On the one hand, the narrator discovers the combination of scenic actions recorded beforehand, based on his own narration of the story. On the other hand, the silent scenes are a support on which he can bring the story to life at each rehearsal or performance, by setting the pace and doubling the virtual stage gestures with his voice. Rehearsals allow us to fine-tune the avatars' scenic intentions with the SIP in line with any changes in the narrative's intentions. In this complex reconstruction of a co-presence between the narrator and the shadow avatars, I direct each partner, physical and virtual, with a view to mutual taming. The awkwardness or stiffness of certain movements is compensated for by the adaptation of the voice. In return, the story of the scholar and his shadow materializes with the mischievous complicity of the shadow avatars.

## 6 Discussion

### 6.1 Deepening programming knowledge

The work of actor and avatar direction that I carried out in the first phase of recording the animations then in the second phase of rehearsing the characters' dialogues was preceded by a programming work in Unreal Engine video game engine. The purpose was to allow recording the acting movements of each avatar and replaying them while being able to suspend the action at key moments. Realized within the AKN_Regie plugin, it is based on the combination of several software within AvatarStaging framework, and the possibilities offered by the video game engine to create idles on the fly without post-production.

The creative process therefore led me to significantly increase my programming skills in Unreal Engine's Blueprint environment to develop the Salient-Idle Player. The development is based on a formalization of theatrical play and the way gestural intentions are organized, which produced convincing results in the show and led to the possibility of fine-tuning pre-recorded stage movements in the direction of autonomy and new intentionality. In turn, this formalization has influenced the way I direct physical actors. In the end, I noticed a personal transformation in relation to the figure of the avatar, and on the avatar's side, the establishment of a simulacrum of autonomy that consolidates the idea of co-species proposed by Couchot.

In terms of acting, the work carried out by the mocaptor enables an intimate confrontation between human physiology and virtual simulation. It is indeed the mocaptor that transmits living movements and presence to the avatar's body. But the avatar's expressive potential offers a vast new territory for the creative imagination. We're not exactly in a situation of real hybridization between a natural human and a virtual human, but we're seeing the necessary modifications to the body schema and proprioception, which pushes the sense of movement towards new developments based on empathy [4]. This work can certainly lead the art of acting in new directions [34] [24].

A the end, the show proposes a simulacrum of an encounter between the spectators and the co-species predicted by Couchot. This prediction is indirectly present in theoretical writings that describe an ongoing process of generalized human avatarization [30] [26]. I'm also following, in Gilbert Simondon's footsteps, the close and necessary articulation between culture and technology: "it is culture, considered as a lived totality, that must incorporate technical ensembles by knowing their nature, in order to be able to regulate human life according to these technical ensembles." [31] I thus personally note the emergence of a new artistic sensibility which is based on my ability to appropriate the computer code to arrange new relations between the various scenic materials, which concern the acting of the actor or the avatar and the sound environment within a mixed reality stage. I am notably struck by the difficulty there is in changing the regime of representations of our physical environment, echoing Walter Ong's reflections on the difficulty of thinking about the regime of an orality without writing, by our minds completely impregnated with written language [27]. How to create now in a digital regime with a mind already impregnated with literacy education?

## 6.2 Perspectives

A first result achieved in *The Shadow* is a simulacrum of a copresence between two types of "living" beings, a narrator, and a group of shadow avatars. It sparked the desire for a true exploration of the autonomy of the avatars through their constraints and potentialities of embryonic existence in their virtual theater. The fictitious existence of the shadow avatars opens new possibilities of coding and acting which would be based on a preliminary development of this autonomy. I begin to take on coding new scores in which the interpretation of actors and avatars is not limited to that of a written text but is based on new emerging expressive tools, irreducible to textual theatrical dramaturgy. In particular, I plan to keep on learning computer programming by reorganizing the existing AKN_Regie blueprint code directly in C++ [25]. This should facilitate the integration of AI algorithmic bricks or specific combinations of existing libraries.

I would like to start with simpler scenic situations than those suggested by Andersen's tale and explore in greater depth the circulation of the shadow avatar's body between 2D and 3D, as well as its texture. As for the mocaptor-avatar relationship, I think that advances in AI could make the process a lot smoother. *The Shadow* is finally the first step towards truly bringing alive artificial shadow avatars switching between 2D and 3D and perhaps striving to friendly coevolve with natural humans.

## ACKNOWLEDGMENTS

We would like to thank didascalie.net which supports the development of the AvatarStaging framework and produces the theatrical performance The Shadow (http://didascalie.net).